\documentclass[twocolumn,notitlepage,preprintnumbers,amsmath,amssymb]{revtex4-1}
\usepackage{graphicx}
\usepackage{bm}
\usepackage[usenames,dvipsnames]{xcolor}

\begin{document}
\preprint{\color{NavyBlue}Preprint \today}
 
\title{\Large \color{NavyBlue}  Remote control of self-assembled microswimmers}

\author{G.Grosjean\footnote{Correspondance : GRASP, Physics Department, University of Li\`ege, B-4000 Li\`ege, Belgium. http://www.grasp-lab.org }}
\author{G.Lagubeau}
\author{A.Darras}
\author{M.Hubert}
\author{G.Lumay}
\author{N.Vandewalle}
\address{GRASP, Institute of Physics B5a, University of Li\`ege, B4000 Li\`ege, Belgium.}

\maketitle


{\bf 
\noindent Physics governing the locomotion of microorganisms and other microsystems is dominated by viscous damping. An effective swimming strategy involves the non-reciprocal and periodic deformations of the considered body. Here, we show that a magnetocapillary-driven self-assembly, composed of three soft ferromagnetic beads, is able to swim along a liquid-air interface when powered by an external magnetic field. More importantly, we demonstrate that trajectories can be fully controlled, opening ways to explore low Reynolds number swimming. This magnetocapillary system spontaneously forms by self-assembly, allowing miniaturization and other possible applications such as cargo transport or solvent flows.
}

~\\
\noindent A Reynolds number much smaller than unity indicates that viscous forces dominate over inertial forces in a given flow. This is usually the case at the microscopic scale, which strongly impacts locomotion mechanisms of both biological and artificial microswimmers. Indeed, to swim by changing its shape, a microscopic body must break time-reversal symmetry \cite{purcell,lauga,powers}. Methods for producing artificial microswimmers are being actively researched \cite{fermigier,aranson,biohybrid}. One major unanswered problem is the control of swimming direction. A fine control of swimming trajectories would be required for most practical applications, such as manipulation and transport of small elements \cite{cargo} or micro-scale fluid flow generation \cite{carpet}. Furthermore, complex fabrication processes could limit potential micro-scale applications. Self-assembled systems present a clear advantage in this regard, as they require no direct manipulation of micro-components \cite{whitesides,pelesko,davies}. Floating soft-ferromagnetic particles have been shown to produce self-assembled structures when exposed to magnetic fields \cite{epje}. Oscillating fields can deform these assemblies in a way that produces low Reynolds locomotion, although by which mechanism remains obscure \cite{sm}. In the present paper, we propose an explanation for the breaking of time-reversal symmetry in magnetocapillary swimmers, as well as demonstrate how swimming speed and direction can be finely controlled in order to produce the desired trajectories.

\begin{figure}
\vskip 0.1 cm
\includegraphics[width=5.0cm]{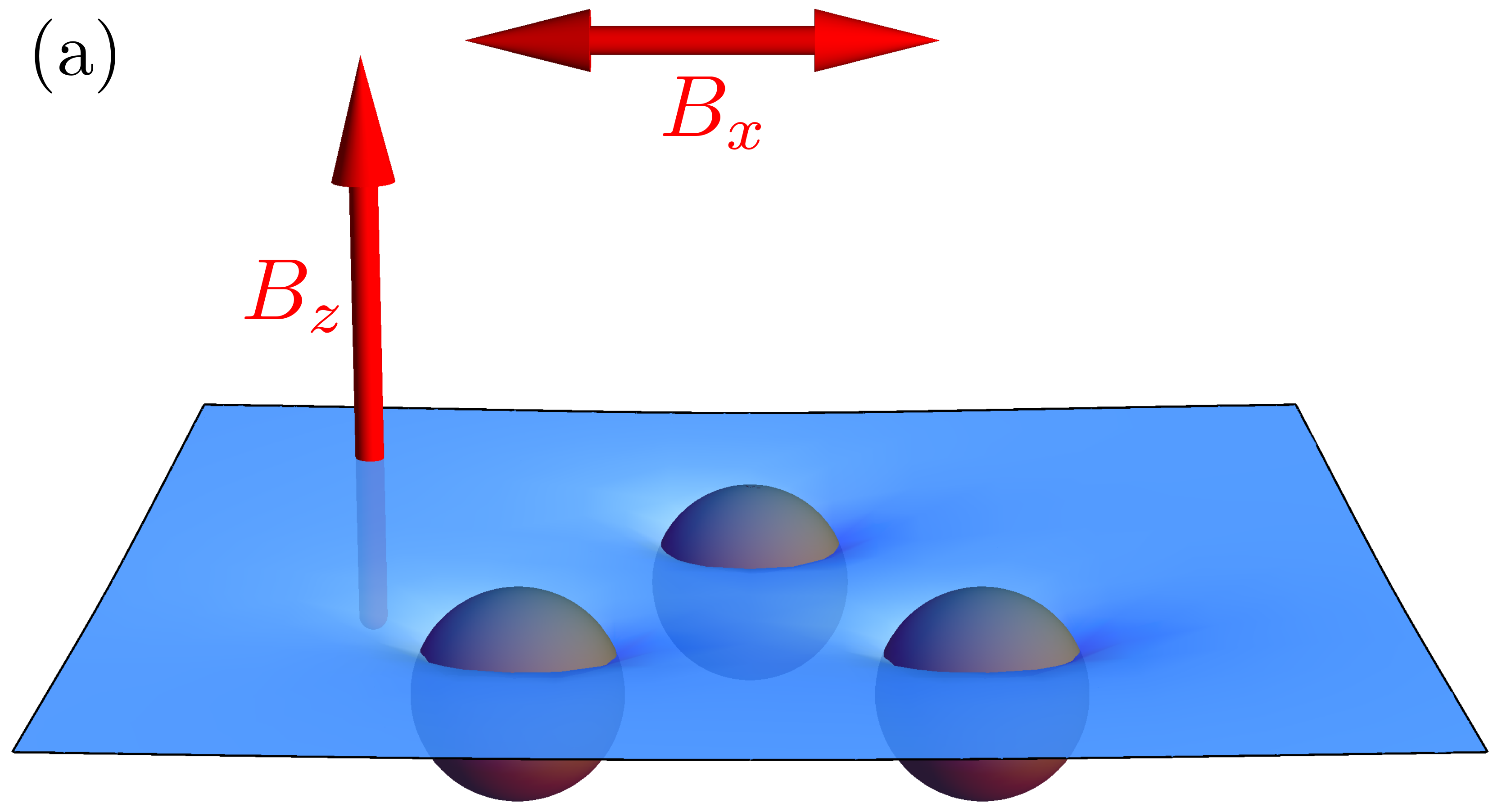}
\includegraphics[width=3.0cm]{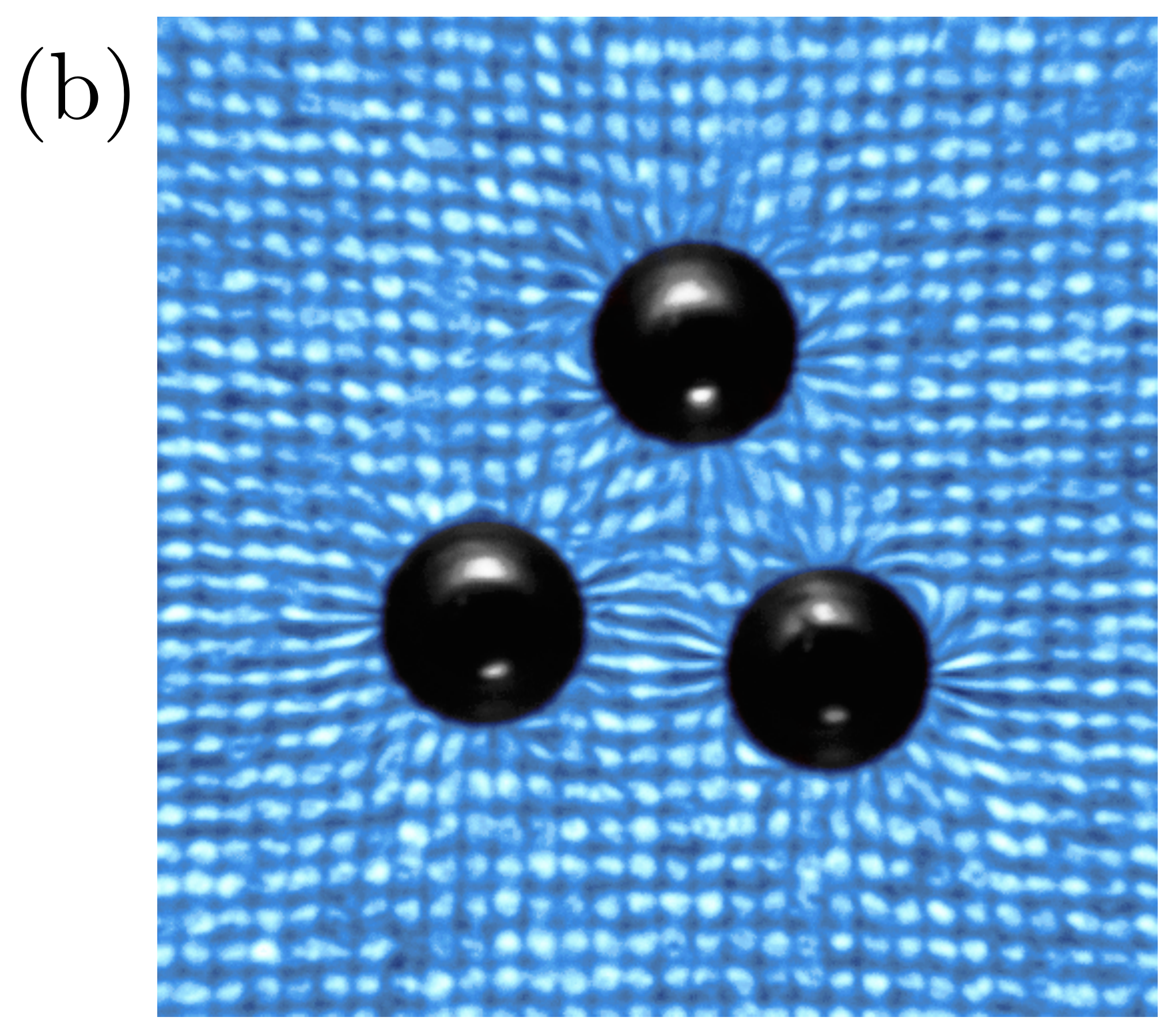}
\vskip 0.2cm
\includegraphics[width=8cm]{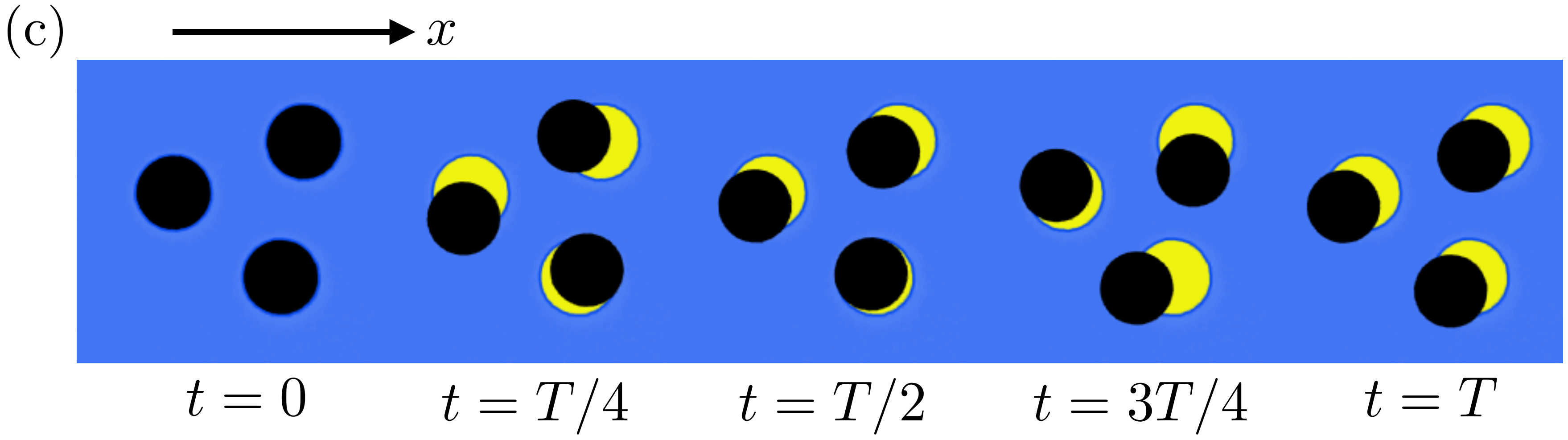}
\vskip -0.2 cm
\caption{{\bf Self-assembled magnetocapillary swimmer} -- (a) Sketch of the magnetocapillary system. Soft ferromagnetic particles are self-assembling at the water-air interface in a Petri dish. A vertical and constant magnetic field $B_z$ is applied through the system. An oscillating horizontal field $B_x$ excites the motion of the self-assembled structure. (b) A top view of the interface emphasizes the self-assembly of $D=500 \, {\rm \mu m}$ beads in a vertical field : capillary attraction is counterbalanced by dipole-dipole repulsion. The deformation of the liquid-air interface is evidenced by placing an array of pixels underneath the Petri dish. (c) Five snapshots of the beads over one period $T$ of the forcing oscillation. The traces of the initial positions are drawn in yellow for emphasizing the motion of the structure. One should notice the rotational oscillation of the structure during the period. }
\label{sketch}
\end{figure}

Three soft ferromagnetic spheres of size $D$ are gently placed along a water-air interface, at the center of a triaxial Helmholtz system. The weight of the particles creates menisci that induce an attractive capillary interaction between neighboring beads \cite{vella}, as shown in Figure \ref{sketch}(a,b). Soft ferromagnetic particles are used, in which magnetic dipoles are induced along the direction of $\vec B$ with negligible hysteretic behavior. A magnetization cycle of the beads can be found at \cite{geoffroy}. As a consequence, the breaking of time reversibility in our system does no come from magnetic hysteresis, as was proposed in recent theoretical models \cite{ogrin}. When the particles are in presence of a magnetic field $\vec B = B_x \vec e_x + B_z \vec e_z$, dipole-dipole interaction competes with capillary attraction. A vertical magnetic field ($B_x = 0$) leads to the formation of a regular triangle, which size can be tuned by $B_z$ \cite{pre,epje}, as shown in the supplementary video 1. In the following, the vertical field is kept constant at $B_z=30 \, {\rm G}$ such that the bead center-to-center distance is close to two bead diameters $2D$. When a constant horizontal field is added, symmetry breaking occurs in the system which then forms an isosceles. An oscillating horizontal field $B_x = \beta_x \sin (2 \pi f t)$ thus periodically deforms the triangular structure. Amplitudes $\beta_x$ larger than $B_z/2$ cannot be reached because the system collapses, leading to hysteretic contacts \cite{pre} between the beads.

\begin{figure}
\vskip 0.1 cm
\includegraphics[width=8.5cm]{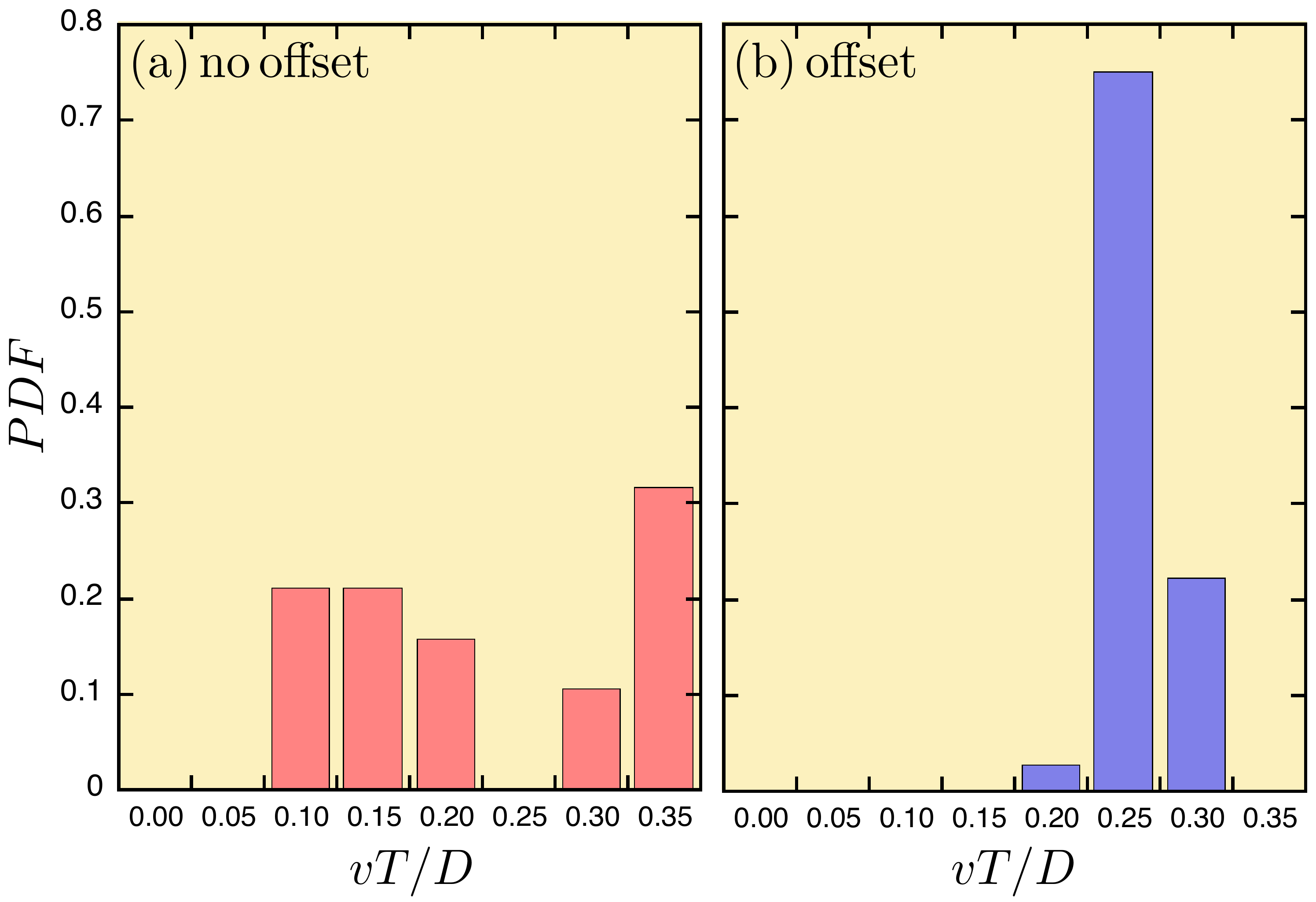}
\vskip -0.2 cm
\caption{{\bf Selection of swimming modes} --Probability Distribution Function (PDF) of normalized speeds obtained in similar  experimental conditions : $\beta_{x}=7.5 \, {\rm G}$ and $f=0.5 \, {\rm Hz}$, but (a) without an offset, (b) with a constant offset $B_{x0}=0.75 \, {\rm G}$. The plots are obtained from 55 independent realizations. }
\label{pdf}
\end{figure}

In Figure \ref{sketch}(c), one observes that small deformations of the triangle are accompanied by large rotational motions of the whole structure. Those repeated deformations/rotations come from the competition between magnetic, capillary and hydrodynamic interactions \cite{sm}. A significant motion of the bead triplet can be seen in Figure \ref{sketch} or in the supplementary video 2. Swimming has been found for frequencies $f$ in between 0.1 Hz and 3 Hz. In the following paper, a frequency of 0.5 Hz has been chosen as a compromise between low Reynolds number and practical constraints linked to experiment running time. The relevant Reynolds number in the system is that of the individual particles, which in this case has typical values around ${\rm Re} \approx 10^{-1}$. It is possible, however, to further lower frequency $f$ and/or field amplitude $\beta_x$ in order to emphasize that low Reynolds locomotion takes place, reaching typical values in the range ${\rm Re} = \left[ 10^{-3}, 10^{-2} \right]$.

~\\
\noindent\textbf{\large\sffamily Results}

\noindent Figure \ref{pdf} presents the Probability Distribution Function (PDF) of the normalized speeds for several swimmers using the same set of parameters : (a) without any offset and (b) with a small offset $B_{x0}= 0.75 \, {\rm G}$. The swimming speed is found to be broadly distributed in the first case while it exhibits a narrow peak in the second case. In fact, various pulsation modes can be observed when the horizontal field oscillates. The offset $B_{x0}$ helps the triplet to select a unique oscillation mode. The horizontal field becomes $B_x= \left[ B_{x0} + \beta_x \sin (2 \pi f t) \right]$. One should remark that the offset enhances the deformation of the triangle but is kept as low as possible here in order to avoid the collapse of the structure. 

\begin{figure}
\vskip 0.1 cm
\includegraphics[width=8 cm]{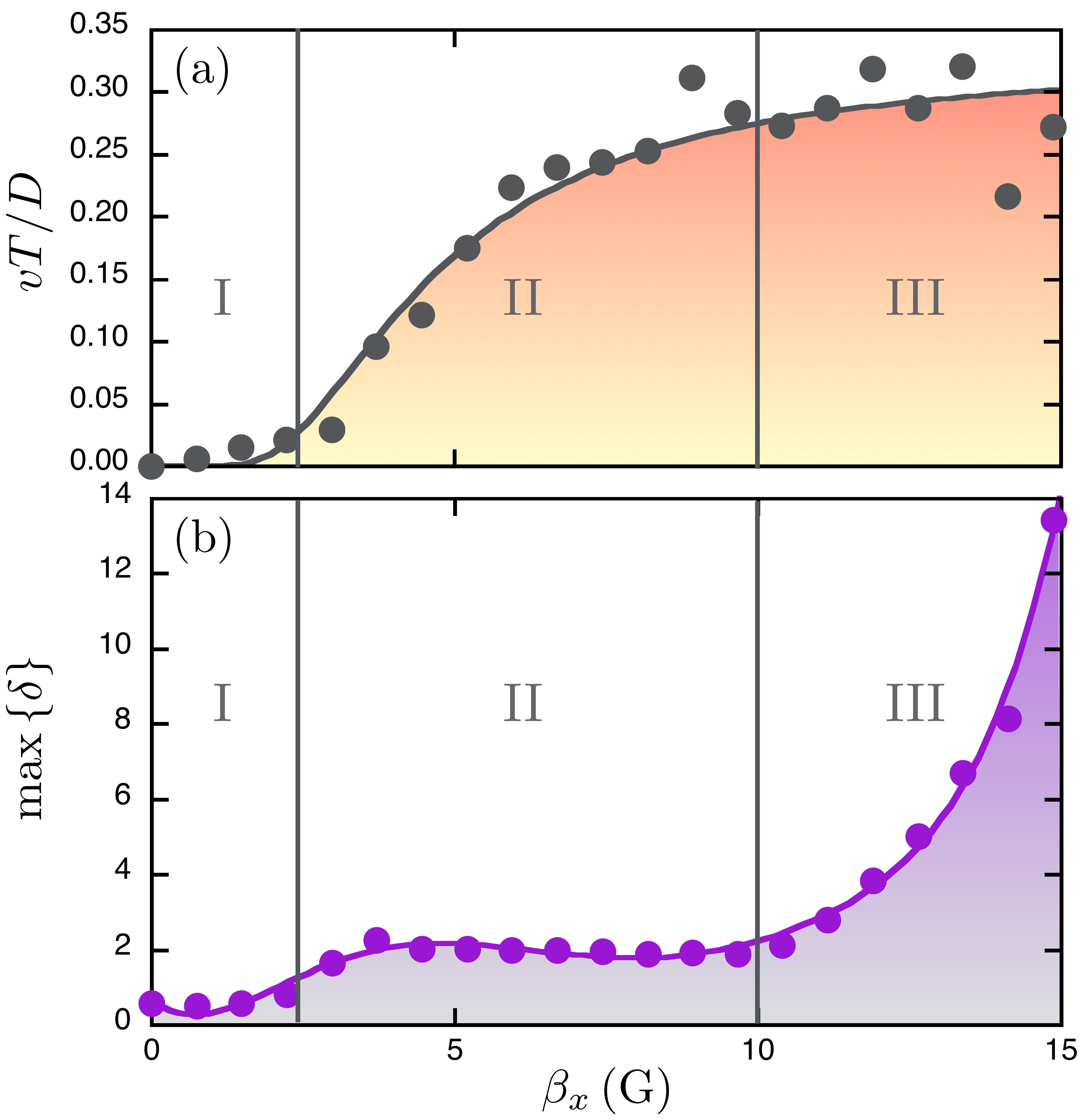}
\vskip -0.2 cm
\caption{{\bf Swimming speeds} -- (a) Average speed of the self-assembled system as a function of the amplitude of oscillations, when the offset is $B_{x0}=0.75 \, {\rm G}$ and $f=0.5 \, {\rm Hz}$. The velocity is seen to increase and saturates around 0.3 $D/T$. The curve is a guide for the eye. (b) The maximum standard deviation $\delta$ of the angles from the regular triangle as a function of the amplitude $\beta_x$. The curve is a guide for the eye. Please note the large differences obtained near the collapse ($\beta_x \approx 15 \, {\rm G}$). Three regimes (I, II and III) are indicated and are discussed in the main text.}
\label{data}
\end{figure}

Figure \ref{data}(a) shows the resulting speed $v$ of the assembly normalized by the bead diameter $D$ per forcing period $T=1/f$, as a function of the amplitude of the field oscillations $\beta_x$. For small oscillations (regime I), no significant motion is observed. When the amplitude $\beta_x$ of the field oscillations becomes much larger than the offset $B_{x0}$ (regime II), the speed increases and saturates at high field values. The dimensionless speed reaches about 0.3 $D/T$, which can be considered an efficient locomotion speed in the Stokes regime \cite{yeomans, spagnolie}. Amplitudes larger than $\beta_x=15 \, {\rm G}$ cannot be used because the system collapses, leading to contacts between the beads. Close to collapse (regime III), speed exhibits large fluctuations, as seen in Figure \ref{data}(a). It has proven hard to control the speed and the pulsation mode in that regime. We also noticed that large offset values reduce the efficient locomotion (regime II) into a sharp range of $\beta_x$ values. 

\begin{figure}
\vskip 0.1 cm
\includegraphics[width=7cm]{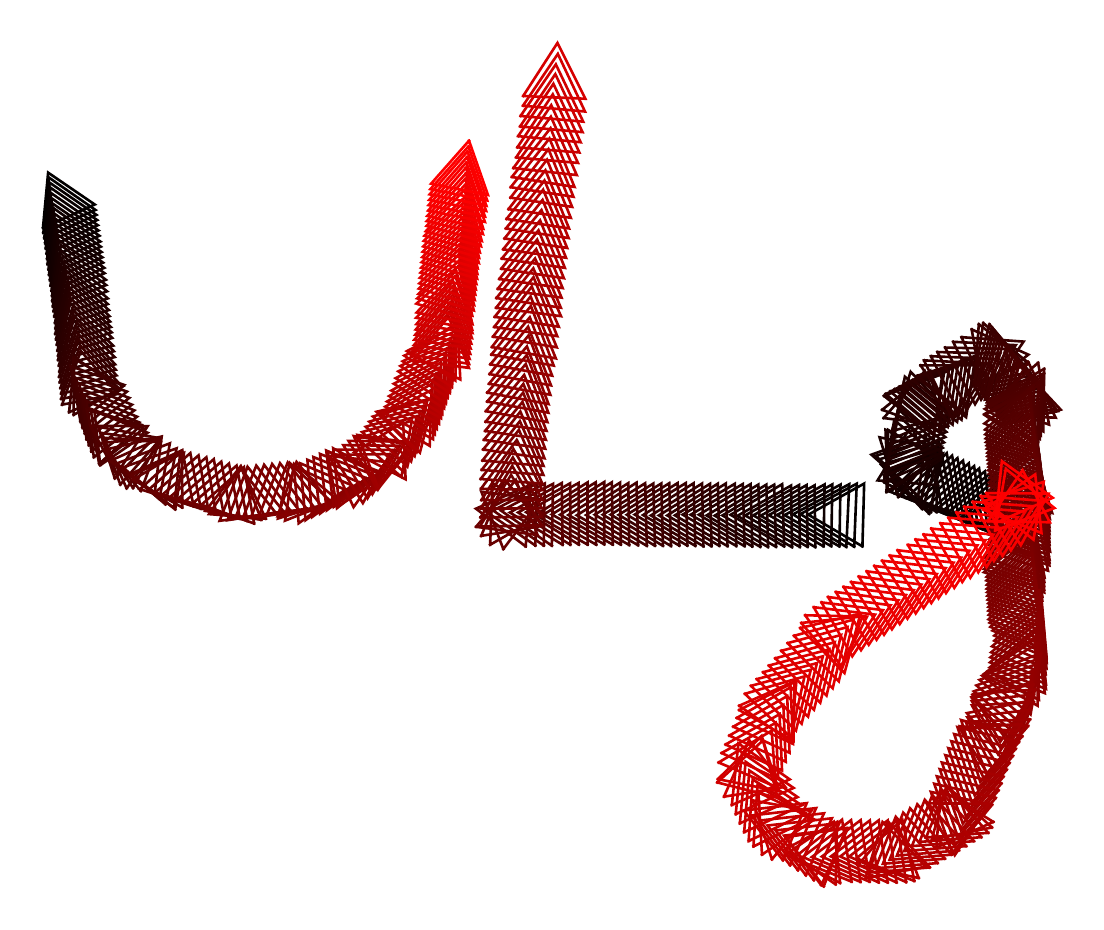}
\vskip -0.2 cm
\caption{{\bf Full control of swimming trajectories} -- The remote control of a microswimmer, by adjusting the horizontal field orientation $\vec e$,  allows us to follow various paths such as U-turn, corner and loops. Laid next to one another, the three letters of ``ULg" are then obtained. Bead centers are tracked for drawing the triangles at each period, emphasizing the orientation of the structure. The colors indicate the time evolution of each trajectory from dark to red. }
\label{ulg}
\end{figure}

In order to quantify the triplet deformations, we computed the standard deviation $\delta$ of the internal angle departures $(\alpha_i-60)$ from the regular triangle (in degrees), i.e. 
\begin{equation}
\delta = \sqrt { {1 \over 3} \sum_{i=1}^3  \left( \alpha_i-60 \right)^2 }\;.
\end{equation}The maximum value of this parameter over a period $\max \left\{ \delta \right\}$ is shown in Figure \ref{data}(b) as a function of $\beta_x$. When $\beta_x$ increases from zero, $\max \left\{ \delta \right\}$ presents a jump similar to $v$ above $B_{x0}$ before saturating to around $2$ degrees (regime II). The speed $v$ and deformation $\delta$ are therefore correlated. It should be stressed that the deformation is relatively weak such that the locomotion should be linked to some additional motion of the beads such as the structure rotation. However, for higher amplitudes close to the collapse (regime III), the triangle is highly stretched providing much higher values of $\delta$ where the speed fluctuates.

The major ingredient triggering the locomotion is the field amplitude $\beta_x$. Once the triangle pulsation is selected, the oscillation mode remains unchanged and the swimmer trajectory is a straight line. Although different from the $x$-axis, the swimming direction is in fact correlated to the oscillating field direction. Indeed, Figure \ref{ulg} presents three independent trajectories of swimmers which are prepared by adjusting slowly the horizontal magnetic field direction in the $x-y$ plane in order to induce U-turn, corner and loops, proving that the remote control can be fully exploited. The paths form the letters of ``ULg", being the logo of our University. The video of the U-turn is given in the supplementary materials. Slowly adjusting the horizontal field orientation offers therefore a convenient way to control the path of a microswimmer.

~\\
\noindent\textbf{\large\sffamily Discussion}

\noindent Two main mechanisms can provide the breaking of time reversibility necessary for low Reynolds locomotion. Symmetry breaking can either occur in the sequence of shapes adopted by the swimmer \cite{purcell} or in the viscous drag felt by the swimmer \cite{tierno}. Here, comparing typical strength of magnetic force $F_{m} \approx \frac{3 \mu_0 m^2}{4 \pi r^4}$ and hydrodynamic interaction $F_{h} \approx \frac{6 \pi \eta R^2 U}{r}$, where $m$ is the magnetic moment of a bead, $R$ its radius, $U$ its speed and $r$ is the center-to-center distance, yields a ratio $\frac{F_{m}}{F_{h}} \approx 5\; 10^2$. The shape of the swimmer is thus driven by magnetic dipole-dipole interactions in our case.

A magnetocapillary model, similar to \cite{spagnolie} and based on the magnetic and capillary interactions only, allows us to capture the possible beads configurations when both vertical and horizontal magnetic fields are applied. Moreover, it provides clues for the origin of the non-reciprocal motion behind low Reynolds locomotion, as seen below. For convenience, distances $r_{ij}$ between beads $i$ and $j$ are adimensionalized by the capillary length, being the characteristic length of capillary interactions $\lambda= \sqrt{\gamma / \rho g} \approx 2.7 \, {\rm mm}$ \cite{vella}.  Considering only quasi-static situations, i.e. excluding hydrodynamic interactions, the dimensionless bead-bead potential is given by 
\begin{equation}
u_{ij}= - K_0(r_{ij})+{{\rm Mc_z} \over r_{ij}^3} + {\rm Mc_x} \left({1\over r_{ij}^3}-{3 (\vec r_{ij} . \vec e_x)^2 \over r_{ij}^5} \right) 
\end{equation}
where $\rm Mc_z$ and $\rm Mc_x$ are magnetocapillary numbers \cite{spagnolie} measuring the competition between magnetic and capillary interactions.  The Bessel function $K_0(r)$ is assumed to capture the attractive capillary interaction \cite{vella}. Each magnetocapillary number is proportional to dipole-dipole interactions. One can fix the number $\rm Mc_z$ in order to obtain an equilibrium distance $r_{ij}=2D/\lambda=0.3$ when the horizontal field is zero ($\rm Mc_{x}=0$, $\rm Mc_z \approx 0.01$). This distance corresponds to what we observe in our experiments. The parameter $\sqrt{\rm Mc_x/Mc_z}$ becomes then the only relevant parameter in the problem, reducing to the ratio $B_x/B_z$. In order to study the equilibrium configurations, we start from a regular triangle with edge length $r=0.3$, the horizontal field is slightly modified and the new equilibrium situation is found by simulated annealing searching for the local minimum of $u= u_{12}+ u_{23}+ u_{31}$, by producing tiny random moves for bead positions. The results of these simulations are discussed below.

\begin{figure}
\includegraphics[width=6.8cm]{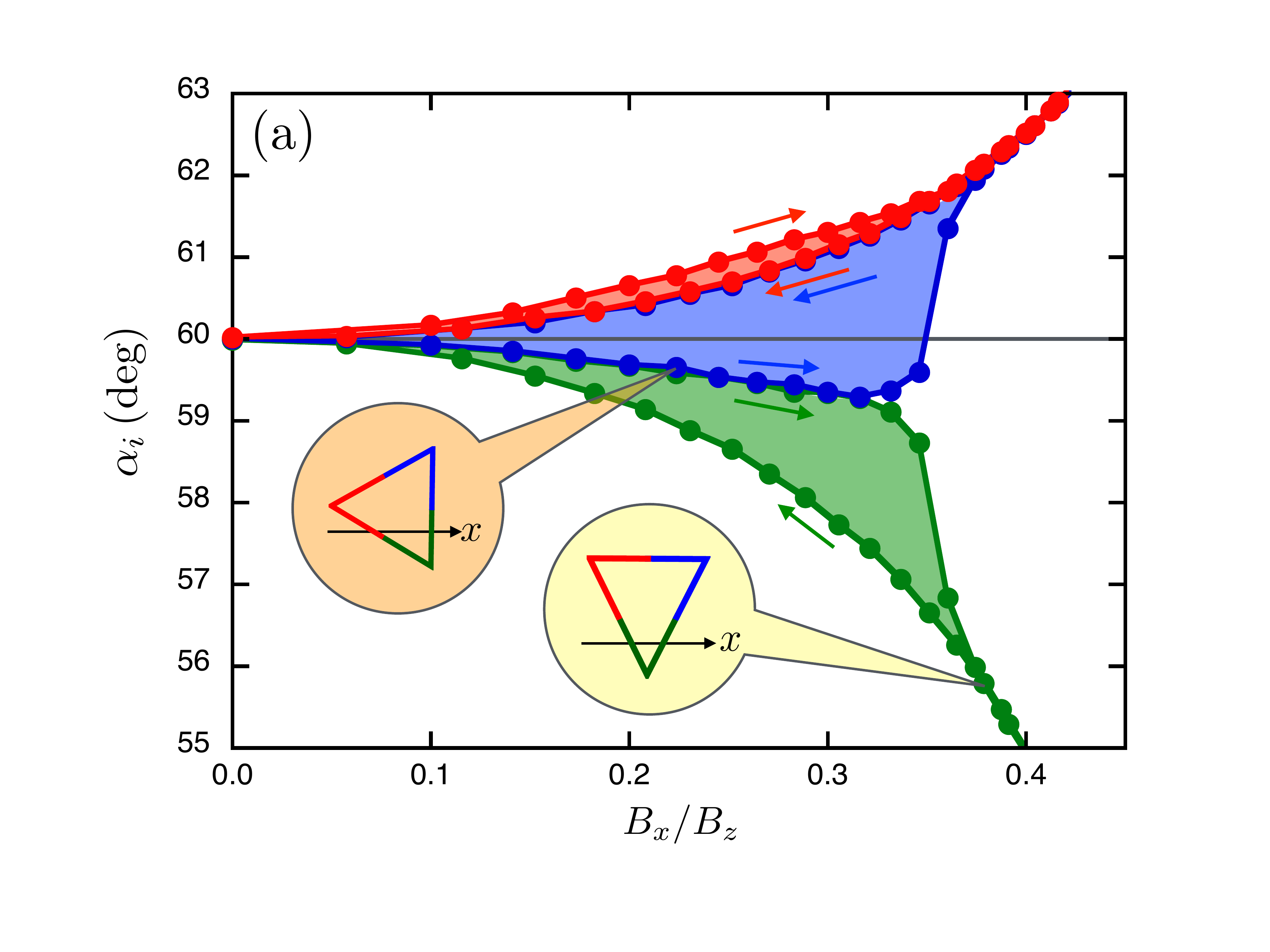}
\includegraphics[width=7cm]{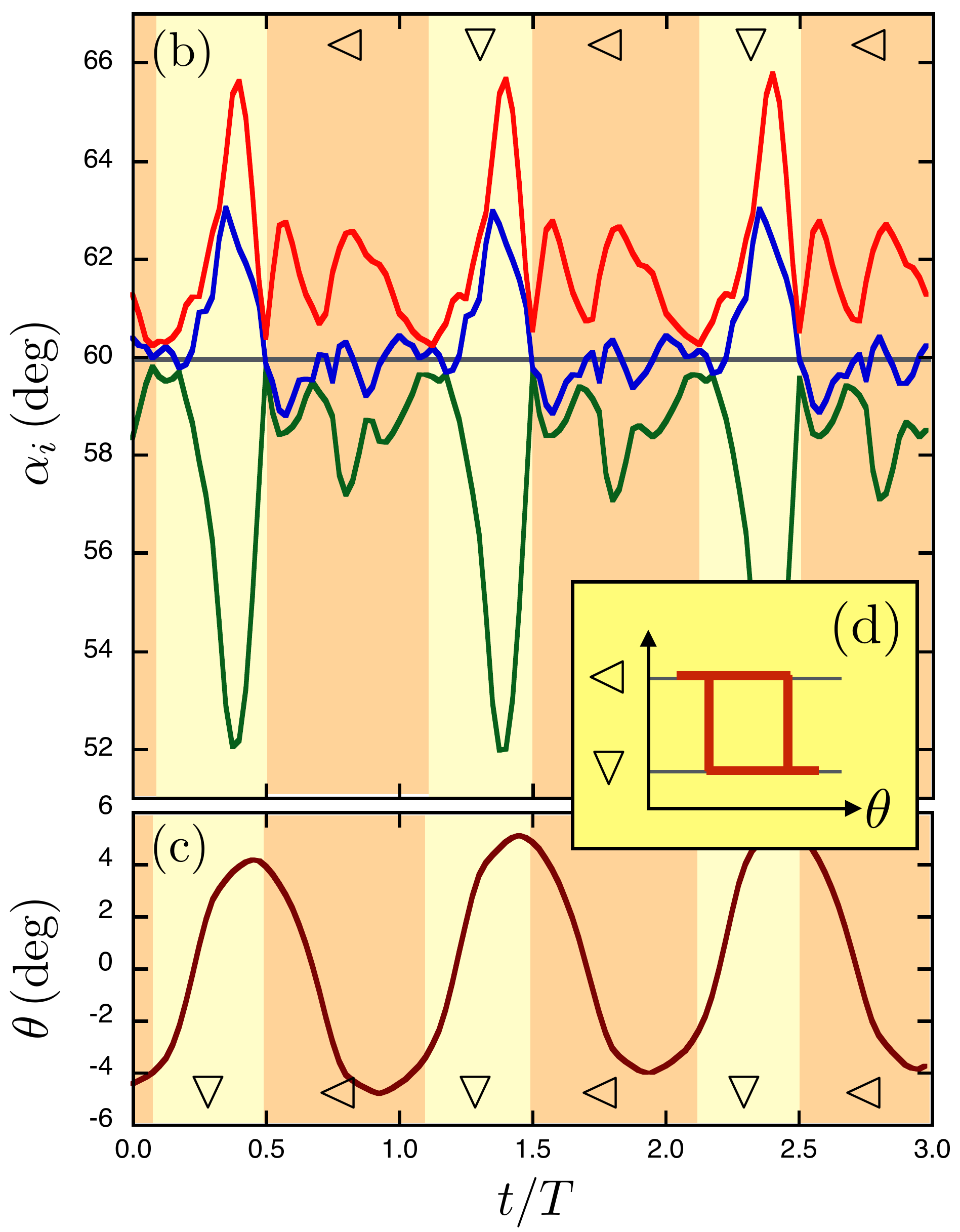}
\vskip -0.2 cm
\caption{{\bf Non-reciprocal motion} -- (a) Internal angles given by the model as a function of the increasing or decreasing horizontal magnetic interaction (see arrows). Angles $\alpha_i$ are colored after being sorted from the smallest to the largest. Two triangles are illustrated using the same color code in order to evidence the orientation of the structure at two specific situations : $\vartriangleleft$ and $\triangledown$ states. (b) Three periods $T$ of the angles $\alpha_i$ using the same color code in the real experiments. One observes that the blue curve oscillates between the red and green ones in each period. The system switches periodically between $\vartriangleleft$ and $\triangledown$ isosceles.  (c) The angle $\theta$ of the whole structure during three periods. Oscillations are seen evidencing the successive switches. (d) This sketch shows the isosceles states as a function of $\theta$ forming a loop at each cycle, i.e. a non-reciprocal motion.}
\label{cycles}
\end{figure}

Figure \ref{cycles}(a) presents the evolution of the internal angles $\alpha_i$ of the triplet at equilibrium and for different values of an increasing then decreasing horizontal field, i.e. when the ratio $B_x/B_z$ follows a quasi-static cycle. Arrows indicate how the structure evolves. The collapse of the structure takes place around $B_x/B_z \approx 1/2$, as expected. In presence of a horizontal field, the triangle is roughly isosceles during the whole cycle since two angles are always close together. At low and increasing field strengths, one angle is larger than $60^\circ$ and the other angles have similar values below $60^\circ$. For the lack of a proper term, we labelled this first type of isosceles ``platy-isosceles". It is characterized by internal angles 
\begin{equation}
\alpha_i (\vartriangleleft)=\left\{60+\sqrt{2} \delta, 60-\frac{\delta}{\sqrt{2}}, 60-\frac{\delta}{\sqrt{2}} \right\}
\end{equation}
and occurs when one edge is perpendicular to the field axis, as sketched in the inset of Figure \ref{cycles}(a). By symmetry, two configurations can be obtained : left ($\vartriangleleft$) and right ($\vartriangleright$).
Above some field strength, the system drastically changes since two angles have similar large values while the later is much smaller than $60^\circ$. One has a second type, labelled ``lepto-isosceles", given by 
\begin{equation}
\alpha_i (\triangledown)=\left\{60+\frac{\delta}{\sqrt{2}}, 60+\frac{\delta}{\sqrt{2}}, 60- \sqrt{2} \delta \right\}
\end{equation}
which is encountered when one side is parallel to the field. Two equivalent configurations are found : up ($\vartriangle$) and down ($\triangledown$). In fact, this bifurcation between platy to lepto-isosceles, e.g. $\vartriangleleft \rightarrow \triangledown$, is accompanied by a rotation of $30^\circ$ of the entire structure, as sketched in the Figure \ref{cycles}(a). The system remains trapped in this last configuration till returning to a regular triangle at zero field. Such a switching behavior between two states is observed periodically in experiments. Indeed, Figure \ref{cycles}(b) presents the evolution of the angles $\alpha_i$ during three successive periods using the same color code : largest angle in red, lowest angle in green. During each period, the main qualitative characteristics predicted by the model are recovered : short successive periods during which both types of isosceles are seen. It should be noted that the model does not predict which configuration is chosen when the field goes back to zero. At zero field, small residual magnetization becomes relevant and might explain why the same configuration is always chosen, such that deformation is periodical. On should also remark that the experimental curves are much more complex than numerical ones due to the presence of hydrodynamic interactions and additional effects, such as a slow rotation of the beads due to tiny residual magnetization. Because of the presence of these additional effects, the angles exhibit periodic jumps from one isosceles type to the other one. Figure \ref{cycles}(c) presents the evolution of the orientation of the entire structure during the same three periods. The angle $\theta$ is seen to oscillate over one period, as expected. Figure \ref{cycles}(d) illustrates the non-reciprocity of the shape succession necessary to propulsion. It has a clear magnetocapillary origin since each magnetic cycle induces a loop in the space of configurations. The successive switches between isosceles states are indeed the driving mechanism of locomotion.

Mesoscale self-assembly \cite{whitesides} is recognized as an elegant way to fabricate microsystems. Because no microfabrication techniques are involved, the experimental setup is quite straightforward. Furthermore, the use of magnetic fields to control the assembly and power the motion offers great flexibility. A magnetocapillary swimmer can indeed be assembled, adjusted, destroyed and reassembled as often as necessary by only changing the magnetic fields. The spontaneous organization of a remote-controlled microswimmer thus represents a valuable achievement. Although the motion of the self-assembled swimmer is constrained along a liquid interface, many applications can be cited : for instance cargo transport, fluid mixing or micromanipulator. Complex functionalities can be reached by considering specific beads and/or a larger number of beads. The fact that the setup does not require the use of chemically active substances is an advantage in this regard. Besides applications, this bead triplet has the great advantage to possess a low number of degrees of freedom such that fundamental aspects of low Reynolds locomotion \cite{olla,najafi} can now be experimentally explored in various conditions.

~\\
\noindent\textbf{\large\sffamily Methods}

\noindent Hereinafter, we present methods and experimental setup. A large Petri dish is filled with water. The liquid/air interface is placed at the center of a triaxial Helmholtz system that compensates the Earth magnetic field, and that is able to produce a uniform field in any direction. When a current $i$ is injected in such coils, a uniform and vertical magnetic field $B_z$ is obtained in the Petri dish. Magnetic fields up to $30 \, {\rm G}$ have been considered. Oscillations of the horizontal field $B_x$ are provided by a function generator and an amplifier. Chrome steel particles (selected alloy AISI 52100, $\rho_s= 7830 \, {\rm kg/m^3}$) are soft ferromagnetic beads, and they do not exhibit any hysteretic behavior in the range of field values used herein. As a result, particles retain a negligible residual magnetic moment once the field is removed. We have estimated this residual magnetization to be roughly 50 times lower than the induced magnetization due to the applied field.  Prior to experiments, spheres are washed with isopropyl alcohol and thereafter dried in an oven.  Different bead diameters have been studied but the results shown herein correspond to $D=500 \, {\rm \mu m}$. At this scale, partial wetting ensures the floatation of the spheres. A CCD camera records images from above. Image analysis provides the position of each bead as a function of time.


~\\
\noindent\textbf{\large\sffamily Acknowledgments}

\noindent This work was financially supported by the FNRS (Grant PDR T.0043.14) and by the University of Li\`ege (Grant FSRC 11/36). GG thanks FRIA for financial support. GLa was financed by the University of Li\`ege and the European Union through MSCA-COFUND-BeIPD project.

~\\
\noindent\textbf{\large\sffamily Author contributions}

\noindent  GG, GLa and AD collected and analyzed experimental data. Physical interpretations were provided by MH, GLu and NV. This manuscript was written by NV. 

~\\
\noindent\textbf{\large\sffamily Additional information}

\noindent \textbf{Supplementary information} accompanies this paper.

\noindent \textbf{Competing financial interests:} The authors declare that they have no competing financial interests.
\\



\end{document}